\documentclass[aps,prd,twocolumn,showpacs,floatfix]{revtex4} 

\usepackage{graphics}
\usepackage{graphicx}

\usepackage{latexsym}
\usepackage[cp866]{inputenc}
\usepackage[english]{babel}

\input epsf

\begin{document}

\title{Interference phenomena in the decay $D^+_s\to\eta\pi^0\pi^+$
induced by \\ the $a^0_0(980)-f_0(980)$ mixing}
\author{N.~N.~Achasov and G.~N.~Shestakov}
\affiliation{Laboratory of Theoretical Physics, S.~L.~Sobolev
Institute for Mathematics, 630090 Novosibirsk, Russia}

\begin{abstract}
Using the data on the decay $D^+_s\to f_0(980)\pi^+\to K^+K^-\pi^+$,
we estimate the amplitude of the process $D^+_s\to\left[f_0(980)\to
(K^+K^-+K^0\bar K^0)\to a^0_0(980)\right]\pi^+\to\eta\pi^0\pi^+$,
caused by the mixing of $a^0_0(980)$ and $f_0(980)$ resonances that
breaks the isotopic invariance due to the $K^+$ and $K^0$ meson mass
difference. Effects of the interference of this amplitude with the
amplitudes of the main mechanisms responsible for the decay $D^+_s
\to\eta \pi^0\pi^+$ are analyzed. As such mechanisms, we examine the
transition $D^+_s\to\eta\rho^+\to\eta\pi^0\pi^+$, which is observed
in experiment, and the possible transition $D^+_s\to\left(a^0_0(980)
\pi^++a^+_0(980)\pi^0\right)\to\eta \pi^0\pi^+$. It is shown that
the rapidly varying phase of the $a^0_0(980)-f_0(980)$ transition
amplitude strongly influences the interference curves.
\end{abstract}


\maketitle


\section{Introduction} \label{Int}

A threshold phenomenon known as the the mixing of $a^0_0(980)$ and
$f_0(980)$ resonances appreciably breaks the isotopic invariance,
since the effect is proportional to $\sqrt{2(M_{K^0}-M_{K^+})/
M_{K^0}}\approx 0.13$ in the modulus of the amplitude \cite{ADS79};
see also Ref. \cite{ADS81}. This effect appears as the narrow (with
the width of about $2(M_{K^0}-M_{K^+})\approx 8$ MeV) resonant peak
between the $K^+K^-$ and $K^0\bar K^0$ thresholds owing to the
transition $a_0^0(980)\to K\bar K\to f_0(980)$ or vice versa
$f_0(980)\to K\bar K\to a^0_0(980)$. There are many proposals in the
literature concerning both the searching of the $a^0_0(980)-f_0(980)
$ mixing and estimating the effects related with this phenomenon;
the detailed list of references may be found, for example, in Ref.
\cite{AKS16}.

Recently, this phenomenon has been discovered experimentally and
studied with the help of detectors VES in Protvino in $\pi^-N$
collisions \cite{Do08,Do11} and BESIII in Beijing in $J/\psi$ decays
\cite{Ab1,Ab2,Ab3}. As a result it has become clear
\cite{AKS15,AKS16,AS16} that the similar isospin breaking effect can
appear not only due to the $a^0_0(980)-f_0(980)$ mixing, but also
due to any mechanism of the production of the $K\bar K$ pairs with
the definite isospin in the $S$ wave, $X\to K\bar K\to
f_0(980)/a^0_0 (980)$ \cite{FN1}. Thus, a new tool emerged to study
the production mechanism and nature of light scalars.


In the present work, we discuss, for the first, time the possibility
of the $a^0_0(980)-f_0(980)$ mixing detection in three-body hadronic
decay of the $D^+_s$ mesons into $\eta\pi^0\pi^+$. We pay attention
to the fact that the manifestation of the isospin-breaking amplitude
$f_0(980)\to K\bar K\to a^0_0(980)$ can be enhanced in this decay
owing to its interference with the amplitudes of other mechanisms.
The sharp and large variation of the phase of the $f_0(980)-
a^0_0(980)$ transition amplitude (by about $90^\circ$ in the region
between $K^+K^-$ and $K^0\bar K^0$ thresholds) plays an important
role in the interference phenomenon. So far, this characteristic
feature of the $a^0_0(980)-f_0(980)$ mixing has remained in the
shadows \cite{FN2,AS04a,AS04b}. By our estimates, the decay
$D^+_s\to\eta \pi^0 \pi^+$ has potential for the $a^0_0(980)-
f_0(980)$ mixing detection.

\section{\boldmath The $a^0_0(980)-f_0(980)$ mixing in $D^+_s\to
\eta\pi^0\pi^+$} \label{SecII}
\subsection{\it{The case of two mechanisms}} \label{SSecIIa}

Figure 1 shows the BaBar data \cite{BaBar11} on the $S$-wave mass
spectrum of the $K^+K^-$ system produced in the decay $D^+_s\to
K^+K^-\pi^+$. Its shape, as well as the shape of the $S$-wave
$\pi^+\pi^-$ spectrum in $D^+_s\to\pi^+\pi^-\pi^+$ \cite{BaBar09},
is approximated by the $f_0(980)$ resonance contribution (see Figs.
1 and 2, and Ref. \cite{FN2a}).

The solid curves in Figs. 1 and 2 are proportional to the modulus
squared of the $f_0(980)$ resonance propagator, i.e.,
$|S_{K^+K^-}|^2\sim1/|D_{f_0}(m^2_{K^+K^-})|^2$, where $m_{K^+K^-}$
is the invariant mass of $K^+K^-$ in the region above the $K^+K^-$
threshold, and $|S_{\pi^+\pi^-}|^2\sim1/|D_{f_0}(m^2_{\pi^+\pi^-}
)|^2$, where $m_{\pi^+\pi^-}$ is the invariant mass of $\pi^+\pi^-$,
respectively. Here the $f_0(980)$ propagator, $1/D_{f_0}$, was taken
from Ref. \cite{AKS16} without any changes.

\begin{figure}
\hspace*{-0.26cm}\includegraphics[width=17pc]{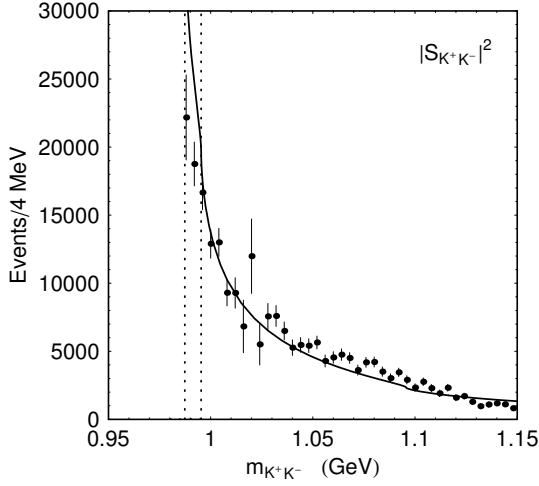}
\caption{\label{Figure1} The BaBar data \cite{BaBar11} on the $S
$-wave $K^+K^-$ mass spectrum in the decay $D^+_s\to K^+K^-\pi^+$.
The data correspond to the modulus squared of the transition
amplitude without the phase space factor of the $K^+K^-$ system in
$D^+_s\to K^+K^-\pi^+$. The dotted vertical lines show the locations
of the $K^+K^-$ and  $K^0\bar K^0$ thresholds. The solid curve
corresponds to the $f_0(980)$ resonance contribution described in
the text.}\end{figure}

\begin{figure}
\hspace*{-0.26cm}\includegraphics[width=17pc]{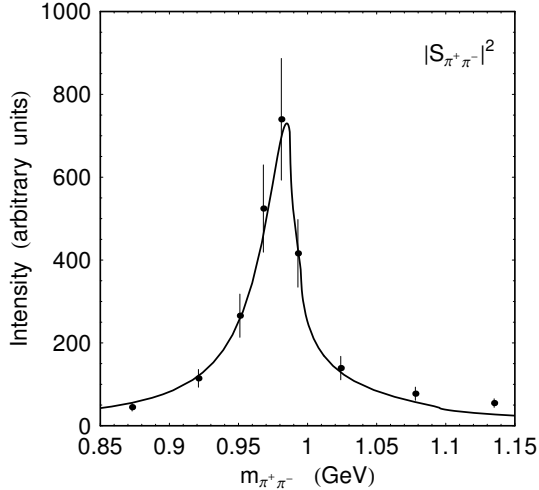}
\caption{\label{Figure2} The BaBar data \cite{BaBar09} on the $S
$-wave $\pi^+\pi^-$ mass spectrum in the decay $D^+_s\to
\pi^+\pi^-\pi^+$; see also Ref. \cite{FN2a}. The shape of the curve
corresponds to the $f_0(980)$ resonance contribution described in
the text.}\end{figure}

The Particle Data Group (PDG) gives \cite{PDG16}
\begin{eqnarray}\label{Eq1}
BR(D^+_s\to f_0(980)\pi^+\to K^+K^-\pi^+) \nonumber \\
=(1.15\pm0.32)\%\,.\qquad\qquad
\end{eqnarray}
This value and its accuracy require further careful study (see
discussions of the assumptions made by BaBar \cite{BaBar11} and CLEO
\cite{CLEO09} with the treatment of the initial data). In fact, in
the original BaBar \cite{BaBar11} and CLEO \cite{CLEO09} analyses a
possible presence of the $a^0_0(980)$ resonance has been neglected
so that the number given in Eq. (\ref{Eq1}) effectively corresponds
to a sum of the $f_0(980)$ and $a_0(980)$ contributions in the
decays of the $D^+_s$ mesons. Therefore, we consider the results of
our analysis as some guide and hope that the detection of the
$a^0_0(980)-f_0(980)$ mixing may shed extra light on the mechanisms
of the $f_0(980)$ and $a^0_0(980)$ production in $D^+_s$ decays.

Using Eq. (\ref{Eq1}), together with the values of the $f_0(980)$
and $a^0_0(980)$ resonance parameters (see Appendix), obtained in
Ref. \cite{AKS16} by analyzing the BESIII data \cite{Ab1} on the
intensity of the $a^0_0(980)-f_0(980)$ mixing in the decays
$J/\psi\to\phi f_0(980) \to\phi a_0(980)\to\phi\eta\pi$ and
$\psi'\to\gamma\chi_{c1}\to \gamma a_0(980)\pi^0\to\gamma
f_0(980)\pi^0\to\gamma \pi^+\pi^-\pi^0 $, we find the following
estimate for the branching ratio of the decay
$D^+_s\to\eta\pi^0\pi^+$ induced by the $a^0_0(980)-f_0(980)$
mixing:
\begin{eqnarray}\label{Eq2}
BR\left(D^+_s\to\left[f_0(980)\to(K^+K^-+K^0\bar K^0)\right.\right.
\nonumber \\ \left.\left.\to a^0_0(980)\right]\pi^+
\to\eta\pi^0\pi^+\right)=4.1\times10^{-4}\,.\ \
\end{eqnarray}
The relevant amplitude of the transition $D^+_s\to\left[f_0(980)
\to(K^+K^-+K^0\bar K^0)\to a^0_0(980)\right]\pi^+\to\eta\pi^0\pi^+$
is presented just below in Eq. (\ref{Eq7}).

The available data on the decay $D^+_s\to\eta\pi^0\pi^+$
\cite{PDG16,CLEO09a,CLEO13} show that it proceeds predominantly via
the $\eta\rho^+$ intermediate state:
\begin{eqnarray}\label{Eq3}
BR(D^+_s\to\eta\rho^+\to\eta\pi^0\pi^+)=(8.9\pm0.8)\%\,,\\
\label{Eq4} BR(D^+_s\to\eta\pi^0\pi^+)=(9.2\pm1.2)\%\,.\qquad
\end{eqnarray}

Let us denote the $D^+_s\to\eta\rho^+\to\eta\pi^0\pi^+$ and
$D^+_s\to\left[f_0(980)\to(K^+K^-+K^0\bar K^0)\to a^0_0(980)
\right]\pi^+\to\eta\pi^0\pi^+$ transition amplitudes as
$A_{\eta\rho^+}$ and $A_{f_0 a^0_0}$, respectively. For the
description of their dependence on the mass variables, we use the
following expressions:
\begin{eqnarray}\label{Eq6}
A_{\eta\rho^+}\equiv A_{\eta\rho^+}(m^2_{\eta\pi^0},m^2_{\eta\pi^+},
m^2_{\pi^0\pi^+})\equiv A_{\eta\rho^+}(s,t,u) \nonumber\\
=C_{D^+_s\eta\rho^+}\,\frac{s-t}{D_{\rho^+}(u)}F_\rho(u)
\sqrt{\frac{g^2_{\rho\pi\pi}}{ 16\pi}}\,,\qquad\qquad\end{eqnarray}
\begin{eqnarray}\label{Eq7}
A_{f_0a^0_0}\equiv A_{f_0a^0_0}(m^2_{\eta\pi^0})\equiv A_{f_0a^0_0}
(s)\qquad\quad\nonumber\\ =C_{D^+_s
f_0\pi^+}\,\frac{\Pi_{a^0_0f_0}(s)}{ D_{a^0_0}(s)
D_{f_0}(s)-\Pi^2_{a^0_0f_0}(s)}\sqrt{\frac{g^2_{a^0_0\eta\pi^0}}{16\pi
}}\,,
\end{eqnarray}
where $s=m^2_{\eta\pi^0}$, $t=m^2_{\eta\pi^+}$, and
$u=m^2_{\pi^0\pi^+}$ are the invariant masses squared of the
indicated meson pairs in the decay $D^+_s\to\eta\pi^0\pi^+$
($\Sigma=s+t+u=m^2_{D_s}+2m^2_\pi+m^2_\eta$, and here we neglect the
$\pi^0$ and $\pi^+$ mass difference and put $m_\pi=0.135$ GeV);
$D_{\rho^+}(u)$, $D_{a^0_0}(s)$, $D_{f_0}(s)$, and $\Pi_{a^0_0
f_0}(s)$ are the inverse propagators of $\rho^+$, $a^0_0(980)$,
$f_0(980)$ resonances and the amplitude of the $a^0_0(980)\to(K^+
K^-+K^0\bar K^0)\to f_0(980)$ transition, respectively, $F_\rho(u)$
is the centrifugal barrier penetration factor (formulas for all
these quantities are presented in Appendix); $g_{\rho\pi\pi}$ and
$g_{a^0_0\eta\pi^0}$ are the coupling constants (see also the
Appendix), $C_{D^+_s\eta\rho^+}$ and $C_{D^+_s f_0\pi^+}$ are the
invariant amplitudes of the decays $D^+_s\to\eta\rho^+$ and
$D^+_s\to f_0(980)\pi^+$, respectively. In so doing, the effective
vertices $D^+_s\to\eta\rho^+$ and $\rho^+\to\pi^0\pi^+$ are taken in
the form
\begin{eqnarray}\label{Eq7a}
V_{D^+_s\eta\rho^+}=
C_{D^+_s\eta\rho^+}(\epsilon^*_{\rho^+},p_{D^+_s}+p_\eta)\,,\quad\\
V_{\rho^+\pi^0\pi^+}=g_{\rho\pi\pi}(\epsilon_{\rho^+},
p_{\pi^+}-p_{\pi^0})\,,\ \ \quad
\end{eqnarray}
where $\epsilon_{\rho^+}$ is the polarization four-vector of the
$\rho^+$ meson, $p_{D^+_s}$, $p_\eta$, $p_{\pi^0}$, and $p_{\pi^+}$
are the four-momenta of the $D^+_s$, $\eta$, $\pi^0$, and $\pi^+$
mesons in the decay $D^+_s\to\eta\pi^0\pi^+$. Hence the kinematical
factor $s-t$ in Eq. (\ref{Eq6}) is $(p_{D^+_s}+p_\eta,p_{\pi^0}-
p_{\pi^+})$. The amplitude $\,A_{ f_0\pi}$, responsible for the
decay $D^+_s\to f_0(980)\pi^+\to K^+K^-\pi^+$ [see Eq. (\ref{Eq1})],
is given by
\begin{eqnarray}\label{Eq7b}
A_{f_0\pi}\equiv A_{f_0\pi}(m^2_{K^+K^-})\equiv A_{f_0\pi}
(s)\qquad\nonumber\\ =C_{D^+_s f_0\pi^+}\,\frac{1}{D_{f_0}(s)
}\sqrt{\frac{g^2_{f_0K^+K^-}}{16\pi }}\,.\qquad
\end{eqnarray}
Each invariant amplitude $C_{D^+_s\eta\rho^+}$ and $C_{D^+_s
f_0\pi^+}$ is represented by two real numbers, a modulus and a
phase, which are independent of the mass variables, i.e.,
$C_{D^+_s\eta\rho^+}=a_1 e^{i\varphi_1}$ and $C_{D^+_s f_0\pi^+}=
a_2e^{i\varphi_2}$. Such an approximation of the amplitudes of heavy
quarkonium decays with the participation of light resonances in
intermediate states is commonly used in the data treatments (fits to
experimental distributions in the Dalitz plots), see, for example,
Refs. \cite{BaBar11,BaBar09,CLEO09}. We use this approximation for
our estimates.

\begin{figure}
\hspace*{-0.26cm}\includegraphics[width=21.5pc]{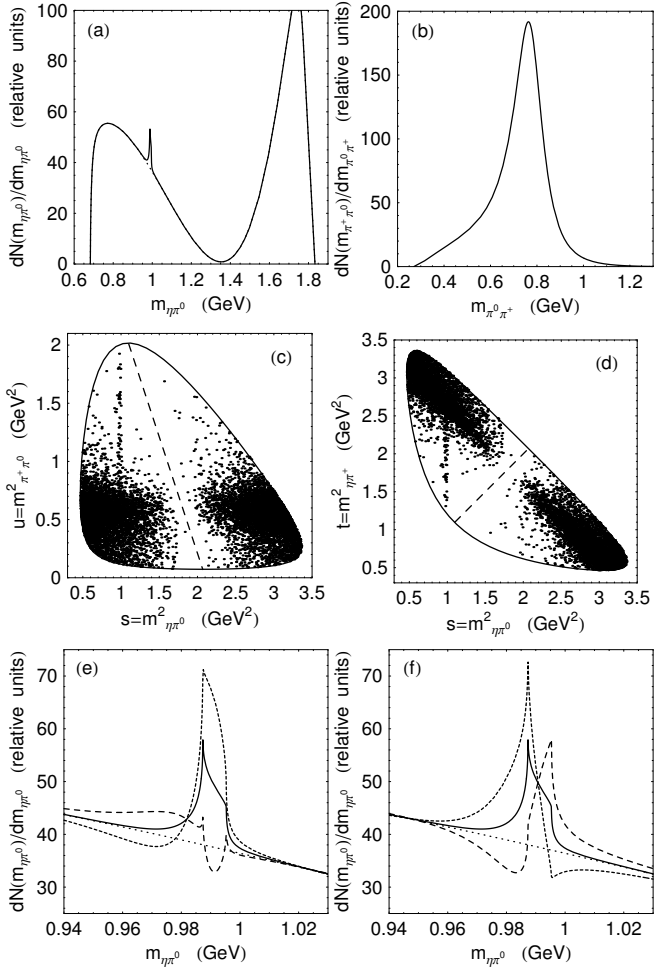}
\caption{\label{Figure3} The illustration of the $a^0_0(980)-f_0
(980)$ mixing manifestation in the decay $D^+_s\to\eta\pi^0\pi^+$
against the mechanism $D^+_s\to\eta\rho^+\to\eta\pi^0\pi^+$. The
solid curves in (a) and (b) show, respectively, the $\eta\pi^0$ and
$\pi^0\pi^+$ mass spectra in the decay $D^+_s\to\eta\pi^0\pi^+$ for
the case of the incoherent sum of the contributions from the
$D^+_s\to\eta\rho^+\to\eta\pi^0\pi^+$ and $D^+_s\to\left[f_0(980)
\to(K^+K^-+K^0\bar K^0)\to a^0_0(980)\right]\pi^+\to\eta\pi^0\pi^+$
mechanisms. The $s$-$u$ and $s$-$t$ Monte Carlo Dalitz plot
distributions for this case are shown in (c) and (d), respectively.
Plots (e) and (f) show the $\eta\pi^0$ mass spectra in the region of
the $K^+K^-$ and $K^0\bar K^0$ thresholds for four variants of the
interference between the amplitudes $D^+_s\to\eta \rho^+\to\eta
\pi^0\pi^+$ and $D^+_s\to\left[f_0(980)\to(K^+K^-+K^0\bar K^0)\to
a^0_0(980)\right]\pi^+\to\eta\pi^0\pi^+$ in comparison with the
incoherent case; the curves are described in the text.}\end{figure}


Taking into account Eqs. (\ref{Eq2}) and (\ref{Eq3}), we present in
Figs. 3(a) and 3(b) the $\eta\pi^0$ and $\pi^0\pi^+$ mass spectra in
the decay $D^+_s\to\eta\pi^0\pi^+$ for the case of the incoherent
sum of the contributions from the $D^+_s\to\eta\rho^+\to\eta\pi^0
\pi^+$ and $D^+_s\to\left[f_0(980)\to(K^+K^-+K^0\bar K^0)\to
a^0_0(980)\right]\pi^+\to\eta\pi^0\pi^+$ mechanisms. The sharp peak
with the width of about $2(m_{K^0}-m_{K^+})\approx8$ MeV in Fig.
3(a) in the region of $K^+K^-$ and $K^0\bar K^0$ thresholds arises
owing to the $a^0_0(980)-f_0(980)$ mixing. Figures 3(c) and 3(d)
show, as an example, the $s$-$u$ and $s$-$t$ Dalitz plots for
approximately $10^4$ $D^+_s\to\eta\pi^0\pi^+$ Monte Carlo events
generated for the above hypothetical case of the incoherent sum of
two mechanisms. As seen from Eq. (\ref{Eq6}), the $s$-$u$ and
$s$-$t$ distributions for the $D^+_s\to\eta\rho^+\to\eta\pi^0\pi^+$
decay mechanism vanish on the dashed lines $u=m^2_{D_s}+2m^2_\pi
+m^2_\eta-2s$\, and \,$t=s$ shown in Figs. 3(c) and 3(d),
respectively. These lines divide the $D^+_s\to\eta\rho^+\to\eta
\pi^0\pi^+$ events into two equal parts. The events caused by the
$a^0_0(980)-f_0 (980)$ mixing concentrate in the vicinity of
$s=m^2_{\eta\pi^0}\approx4m^2_K$ on the $s$-$u$ and $s$-$t$ Dalitz
plots. They make up about one-hundredth of a half of the
$D^+_s\to\eta\rho^+\to\eta\pi^0 \pi^+$ events [see Eqs. (\ref{Eq2})
and (\ref{Eq3})]. This is large for the isospin breaking
contribution which, at the first sight, could be naturally expected
to have the magnitude at the level of $(m_d-m_u)/\bar m$ (where
$m_d$, $m_u$, $\bar m=(m_d+m_u)/2$ are the constituent-quark masses)
or $\alpha=e^2/4\pi$ (electromagnetic constant) in the reaction
amplitude and thus at the level of $10^{-4}$ in the amplitude
squared.

Figures 3(e) and 3(f) show four variants of the $\eta\pi^0$ mass
spectrum in the region of the $K^+K^-$ and $K^0\bar K^0$ thresholds
with taking into account the interference of the $D^+_s\to\eta
\rho^+\to\eta\pi^0\pi^+$ transition amplitude $A_{\eta\rho^+}$ and
the amplitude $A_{f_0a^0_0}$ caused by the $a^0_0(980)-f_0(980)$
mixing,
\begin{equation}\label{Eq5}
\frac{dN_{\eta\pi^0}}{dm_{\eta\pi^0}}=\int\limits_{a_-(s)}^{a_+(s)}
|A_{\eta\rho^+}+A_{f_0a^0_0}|^2\,2m_{\eta\pi^0}dm^2_{\pi^0\pi^+}\,.
\end{equation}
Here the integration is made over the physical region of the
variable $m^2_{\pi^0\pi^+}=u$ from $a_-(s=m^2_{\eta\pi^0})$ to
$a_+(s=m^2_{\eta\pi^0})$, where
\begin{eqnarray}\label{EqIV-1}
a_\pm(s)=\frac{1}{2}\left(\Sigma-s-\frac{(m^2_{D_s}-m^2_\pi)(m^2_\eta
-m^2_\pi)}{s}\right)\qquad\nonumber\\
\pm\frac{2m_{D_s}}{\sqrt{s}}\,p(s)q(s)\,,\quad\qquad \\
p(s)=\sqrt{m^4_{D_s}-2m^2_{D_s}(s+m^2_\pi)+(s-m^2_\pi)^2}/(2m_{D_s}),\
\ \\ q(s)=\sqrt{s^2-2s(m^2_\eta+m^2_\pi)+(m^2_\eta-m^2_\pi)^2}
/(2\sqrt{s}).\ \ \ \ \end{eqnarray} Using the data from Eqs.
(\ref{Eq1}) and (\ref{Eq3}), we find $C_{D^+_s f_0\pi^+}/C_{D^+_s
\eta\rho^+}=(a_2/a_1)\,\xi \approx(4.5\,\,\mbox{GeV})\,\xi$, where
$\xi=e^{i\varphi_{21}}$ and $\varphi_{21}=\varphi_2-\varphi_1$ is
the relative phase of the amplitudes $C_{D^+_s f_0\pi^+}$ and
$C_{D^+_s \eta\rho^+}$. This phase is unknown and to illustrate the
possible interference patterns we put $\varphi_{21}$\,=\,$0^\circ$,
$\pm90^\circ$, and $180^\circ$ (respectively, $\xi\,=\,1$, $\pm i$,
and $-1$). The short and long dashed curves in Fig. 3(e) show the
$\eta\pi^0$ mass spectra for $\xi=1$ and $\xi=-1$, respectively. The
dotted curve in this figure shows the contribution from the
amplitude $A_{\eta\rho^+}$ only, and the solid curve corresponds to
the above case of the incoherent sum of two mechanisms. The solid
and dotted curves in Fig. 3(f) show the same as in Fig. 3(e), and
the short and long dash curves illustrate the interference patterns
corresponding to $\xi=i$ and $\xi=-i$, respectively.

Note that the interference of $A_{f_0a^0_0}$ with the other
contributions will be practically always essential (see Figs. 3(e)
and 3(f)) in consequence of the sharp change of the phase of the
$a^0_0(980)-f_0(980)$ transition amplitude $\Pi_{a^0_0f_0}(s)$ by
about $90^\circ$ in the region between $K^+K^-$ and $K^0\bar K^0$
thresholds \cite{AS04a, AS04b,AKS16}, where the modulus of
$\Pi_{a^0_0f_0}(s)$ is maximal and approximately constant (see
Appendix for details).

\subsection{\it{The case of three mechanisms}} \label{SSecIIb}

In principle, the decay $D^+_s\to\eta\pi^0\pi^+$ can proceed not
only via the $\eta\rho^+$ intermediate state but also via the
$(a_0(980)\pi)^+$ production, $D^+_s\to[a^+_0(980)\pi^0+
a^0_0(980)\pi^+]\to\eta\pi^0\pi^+$. However, such a transition
should be expected to be small. Based on the data quoted in Eqs.
(\ref{Eq3}) and (\ref{Eq4}), we put $BR(D^+_s\to(a_0(980)\pi)^+\to
\eta \pi^0\pi^+)\approx1\%$ as a very rough upper estimate. Note
that by our estimate the relevant upper limit for $BR(D^+_s\to
a^0_0(980)\pi^+\to K^+K^-\pi^+)$ is $\approx0.1\%$. This estimate
consists with the initial dominance of the $f_0(980)$ resonance in
the decay $D^+_s\to f_0(980)\pi^+\to K^+K^-\pi^+$ (see Eq. (1) and
the discussion after it, and also Ref. \cite{FN3}).

Thus, we have three interfering mechanisms of the decay
$D^+_s\to\eta\pi^0 \pi^+$. The corresponding decay amplitude is
\begin{equation}\label{Eq8}
A_{D^+_s\to\eta\pi^0\pi^+}=A_{\eta\rho^+}+A_{ f_0a^0_0}+ A_{a_0\pi
}\,, \end{equation} where the amplitude $A_{a_0\pi}$ describes the
transition $D^+_s\to(a_0(980)\pi)^+\to\eta\pi^0\pi^+$. Like
$A_{\eta\rho^+}$ [see Eq. (\ref{Eq6})], the amplitude $A_{a_0\pi}$
has to be antisymmetric with respect to permutation of the $s$ and
$t$ variables \cite{FN4}. Taking this into account, we approximate
the amplitude $A_{a_0\pi}$ by the following expression
\begin{eqnarray}\label{Eq9} A_{a_0\pi}\equiv A_{a_0\pi}(m^2_{\eta
\pi^0},m^2_{\eta\pi^+})\equiv A_{a_0\pi}(s,t)\qquad\nonumber\\
=C_{D^+_sa^0_0\pi^+}\,\left[\frac{1}{D_{a^0_0}(s)}-\frac{1}{D_{a^+_0}
(t)}\right]\sqrt{\frac{g^2_{a_0\eta\pi^0}}{16\pi}}\,,\quad
\end{eqnarray}
where the production amplitude $C_{D^+_sa^0_0\pi^+}=a_3e^{i
\varphi_3}$ is assumed to be the $s$- and $t$-independent complex
constant. Note that any coherent sum of the amplitudes $A_{\eta
\rho^+}$ and $A_{a_0\pi}$ gives the symmetric distribution of the
$\eta\pi^0\pi^+$ events in the $s-t$ Dalitz plot relative to the
$t=s$ line. The isospin-breaking amplitude $A_{f_0a^0_0}=A_{f_0
a^0_0}(s)$ caused by the $a^0_0(980)-f_0(980)$ mixing depends
exclusively on $s$ and therefore is responsible for the asymmetry of
the distribution of the $\eta\pi^0\pi^+$ events in the $s-t$ Dalitz
plot (relative to the $t=s$ line).

By our estimate $C_{D^+_sa^0_0\pi^+}/C_{ D^+_s\eta\rho^+}=(a_3/a_1)
\xi'\approx(1.65\,\,\mbox{GeV})\,\xi'$, where $\xi'=e^{i\varphi_{31
}}$ and $\varphi_{31}=\varphi_3-\varphi_1$ is an unknown relative
phase of the amplitudes $C_{D^+_sa^0_0\pi^+}$ and $C_{D^+_s\eta
\rho^+}$. We examined 16 variants of the interference patterns
corresponding to different combinations of the relative phase values
$\varphi_{21}$\,=\,$0^\circ$, $\pm90^\circ$, $180^\circ$ and
$\varphi_{31}$\,=\,$0^\circ$, $\pm90^\circ$, $180^\circ$ or
parameters $\xi=1$, $\pm i$, $-1$ and $\xi'=1$, $\pm i$, $-1$. To
illustrate possible manifestations of the $a^0_0(980)-f_0(980)$
mixing effect, we chose 4 of them with $(\xi,\xi')=(i,-1)$,
$(-1,1)$, $(1,i)$, and $(1,-1)$. The solid curves in Figs. 4(a) and
4(c) show the $\eta\pi^0$ mass spectra,
\begin{equation}
\label{Eq10}\frac{dN_{\eta\pi^0}}{dm_{\eta\pi^0}}=\int\limits_{a_-
(s)}^{a_+(s)}|A_{D^+_s\to\eta\pi^0\pi^+}|^22m_{\eta\pi^0}dm^2_{
\pi^0\pi^+}\,,\end{equation} calculated with the use of Eqs. (5),
(6), (11)--(15). The corresponding distributions of the Monte Carlo
events ($\sim |A_{D^+_s\to\eta \pi^0\pi^+}|^2$) in the $s$-$t$
Dalitz plots are shown in Figs. 4(b) and 4(d). The variant
represented in Figs. 4(a) and 4(b) corresponds to combination $(\xi,
\xi')=(i,-1)$ for which the influence of the $a^0_0(980)-f_0(980)$
mixing seems most appreciable. The variant represented in Figs. 4(c)
and 4(d) corresponds to combination $(\xi,\xi')=(-1,1)$. In this
case, the $\eta\pi^0$ mass spectrum demonstrates a small narrow peak
located on the smooth background in the region of the $K\bar K $
thresholds [see Fig. 4(c)]. Nevertheless, the asymmetry effect is
clearly visible in the Dalitz plot [see Fig. 4(d)] (though it almost
collapses in the $\eta\pi^0$ projection). The mass spectra
$dN_{\eta\pi^0}/dm_{\eta\pi^0}$ in the $a^0_0(980)$ resonance region
are presented in more detail in Fig 5 for variants with
$(\xi,\xi')=(i,-1)$, $(-1,1)$, $(1,i)$, and $(1,-1)$. The dotted
curves in Figs. 4(a), 4(c), and 5 correspond to the mass spectra
$dN_{\eta\pi^0}/dm_{\eta\pi^0}$ without the contribution of the
amplitude $A_{f_0a^0_0}$. Note that the asymmetry in the $s$-$t$
Dalitz plot distributions relative to the $t=s$ line (see Fig. 4)
manifests itself in all considered 16 variants.

\begin{figure}
\hspace*{-0.5cm}\includegraphics[width=22.pc]{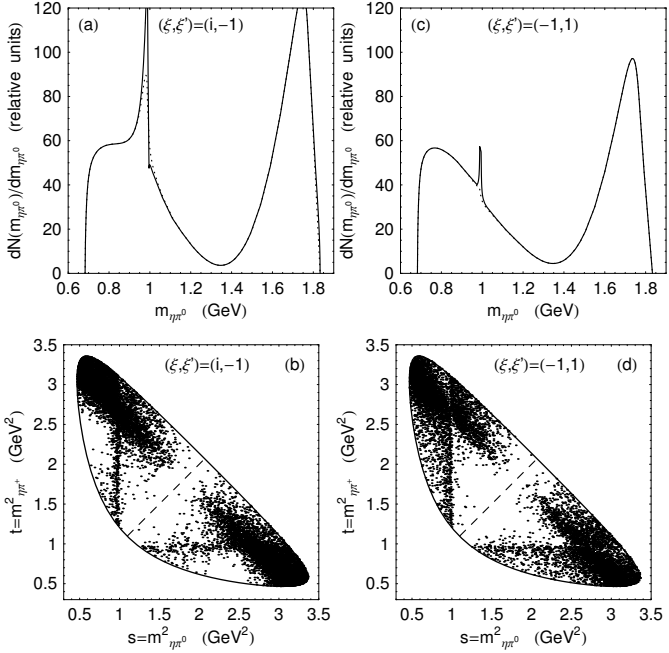}
\caption{\label{Figure5} The illustration of possible manifestations
of the $a^0_0(980)-f_0(980)$ mixing effect in the decay $D^+_s\to
\eta\pi^0\pi^+$ for the case of three interfering mechanisms. The
solid curves in (a) and (c) show the mass spectra $dN_{\eta\pi^0}/
dm_{\eta\pi^0}$ calculated with the use of Eqs. (\ref{Eq8}) and
(\ref{Eq10}) for two sets of the relative phases indicated in the
plots. The corresponding $s$-$t$ Monte Carlo Dalitz plot
distributions ($\sim |A_{D^+_s\to\eta \pi^0\pi^+}|^2$) are shown in
(b) and (d). The mass spectra without the contribution of the
amplitude $A_{f_0a^0_0}$ are shown in (a) and (c) by the dotted
curves (see Fig. 5 for details).}\end{figure}

\begin{figure}
\hspace*{-0.5cm}\includegraphics[width=22.pc]{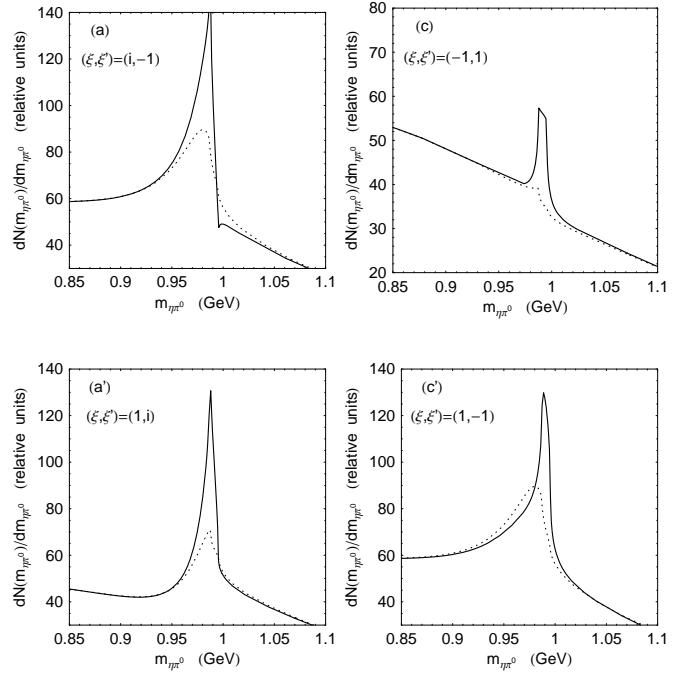}
\caption{\label{Figure7} The solid curves in plots (a), (c) and
(a$'$), (c$'$) show the $\eta\pi^0$ mass spectra in the $a^0_0(980)$
resonance region corresponding to the interference variants with
$(\xi,\xi')=(i,-1)$, $(-1,1)$ [see Figs. 4(a) and 4(c)] and $(1,i)$,
$(1,-1)$, respectively. The dotted curves correspond the mass
spectra without the contribution of the amplitude $A_{f_0a^0_0}$.}
\end{figure}

Detecting signs of the $D^+_s\to[a^0_0(980)\pi^++a^+_0(980)
\pi^0]\to\eta\pi^0\pi^+$ decay mechanisms is one of the interesting
problems both for the weak hadronic decay physics of the $D^+_s$
meson and for the physics of the light scalar $a_0(980)$ and
$f_0(980)$ mesons. At present, intensive investigations in these
lines are realized by the LHCb, BaBar, CLEO, Belle, and BESIII
Collaborations (see, for example, recent reviews
\cite{PDG16,Nogu15,Reis16,Lois16}).

\section{Conclusion and discussion} \label{SecIII}

Light meson spectroscopy from hadronic charm meson decays (in
particular, study of the $a^0_0(980)$ and $f_0(980)$ resonances) is
one of the main lines of the LHCb program on charm physics
\cite{Nogu15,Reis16}. It is hoped that the measurements of the
$D^+_s$ meson decays with huge statistics, really reachable at LHCb,
will allow us to reveal the isospin breaking effect caused by the
$a^0_0(980)-f_0(980)$ mixing in the $D^+_s\to\eta\pi^0\pi^+$ channel
and obtain new information on the production mechanisms and nature
of the light scalar mesons.

Note that the investigations of the $a^0_0(980)-f_0(980)$ mixing in
three-body decays of the $D^0$ meson, such as $D^0\to
K^0_S\pi^+\pi^-$, $D^0\to K^0_S\eta\pi^0$, $D^0\to\bar K^0K^-K^+$,
$D^0\to K^-K^+\pi^0 $, and $D^0\to\pi^+\pi^-\pi^0$, are also
promising and interesting. These decays differ appreciably from
those of the $D^+_s$ meson. We hope to present detailed estimates
for the case of the $D^0$ decays elsewhere in the near future.

Note also that the $a^0_0(980)-f_0(980)$ mixing in the semileptonic
decays $D^+_s\to[\pi^0\eta,\pi\pi]e^+\nu$ has been discussed
recently in Ref. \cite{Wa16}.


\begin{center}\vspace*{0.25cm}{\small\bf ACKNOWLEDGMENTS}\end{center}
The present work is partially supported by the Russian Foundation
for Basic Research Grant No. 16-02-00065 and the Presidium of the
Russian Academy of Sciences Project No. 0314-2015-0011.


\begin{center}{\small\bf\boldmath APPENDIX: PROPAGATORS AND $a^0_0(980)-f_0
(980)$ MIXING AMPLITUDE}\end{center}

The inverse propagator of the $\rho^+$ meson in Eq. (\ref{Eq6}) is
\begin{equation}\label{11}
D_{\rho^+}(u)=m^2_\rho-u-i\,\sqrt{u}\,\Gamma_\rho(u)\,,
\end{equation} where
$\Gamma_\rho(u)=(m^2_\rho/u)\Gamma_\rho[q(u)/q(m^2_\rho)]^3F^2_\rho(u)$,
$F^2_\rho(u)=[1+r^2_\rho q^2(m^2_\rho)]/[1+r^2_\rho q^2(u)]$,
$r_\rho=5$ GeV$^{-1}$, $q(u)=\sqrt{u-4m^2_\pi}/2\,$, $m_\rho=0.775$
GeV, $\Gamma_\rho=0.148$ GeV, $g^2_{\rho\pi\pi}/(4\pi)=2.8$
\cite{PDG16}.

The $a^0_0(980)-f_0(980)$ mixing amplitude in Eq. (\ref{Eq7}),
caused by the diagrams shown in Fig. \ref{Figure1}, has the form
\begin{figure}[!ht]
\hspace*{0.3cm}\includegraphics[width=17pc]{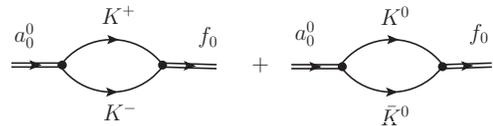}
\caption{\label{Figure1} The $K\bar K$ loop mechanism of the
$a^0_0(980)-f_0(980)$ mixing.}\end{figure}
\begin{eqnarray}\label{Eq12} \hspace*{-35pt} &&
\Pi_{a^0_0f_0}(s)=\frac{g_{a^0_0K^+K^-}g_{f_0K^+K^-
}}{16\pi}\Biggl[\,i\,\Bigl(\rho_{K^+K^-}(s)\nonumber\\
\hspace*{-35pt} && -\rho_{K^0\bar K^0}(s)\Bigr)-
\frac{\rho_{K^+K^-}(s)}{\pi}\ln\frac{1+\rho_{K^+K^-}(s)}
{1-\rho_{K^+K^-}(s)}\nonumber\\ \hspace*{-35pt} && +\frac{\rho_{K^0
\bar K^0}(s)}{\pi}\ln\frac{1+\rho_{K^0\bar K^0}(s)}{1-\rho_{K^0\bar
K^0}(s)}\,\,\Biggl]\,,\end{eqnarray} where $s$ (the square of the
invariant virtual mass of scalar resonances) $\geq4m^2_{K^0}$ and
$\rho_{K\bar K}(s)=\sqrt{1-4m_K^2/s}$; in the region $0\leq
s\leq4m^2_K$, $\rho_{K\bar K}(s)$ should be replaced by
$i|\rho_{K\bar K}(s)|$. The modulus and the phase of $\Pi_{a^0_0f_0}
(s)$ are shown in Fig. \ref{Figure2ab}. Since $\Pi_{a^0_0f_0}(s)$ is
not small between the $K\bar K$ thresholds, all orders of the
$a^0_0(980)-f_0(980)$ mixing has been taken into account in Eq.
(\ref{Eq7}) for the amplitude $A_{f_0a_0}$ \cite{ADS79,ADS81,
AKS16}.
\begin{figure}[!ht] \hspace{-0.35cm}
\includegraphics[width=20pc]{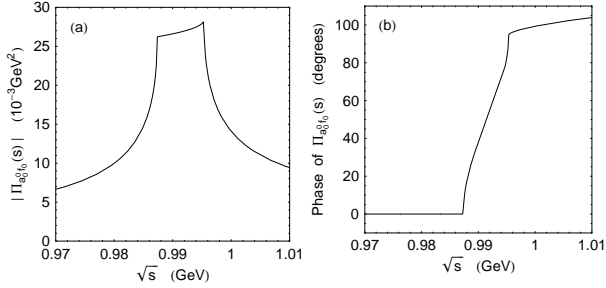}
\caption{\label{Figure2ab} (a) An example of the modulus of the
$a^0_0(980)-f_0(980)$ mixing amplitude. (b) The phase of the
$a^0_0(980)-f_0(980)$ mixing amplitude.}\end{figure}

In Eq. (\ref{Eq7}), $D_r(s)$ is the inverse propagator of the
unmixed resonance $r$ $[r=a^0_0(980),f_0 (980)]$ with the mass
$m_r$\,, \begin{eqnarray}\label{Eq13}
D_r(s)=m^2_r-s+\sum_{ab}[\mbox{Re}\Pi^{ab}_r(m^2_r)-\Pi^{ab}_r(s)],
\end{eqnarray} $ab=(\eta\pi^0,\,K^+K^-,\,K^0\bar K^0,\,\eta'\pi^0)$
for $r=a^0_0(980)$ and $ab=(\pi^+\pi^-,\,\pi^0\pi^0,\,K^+K^-,\,K^0
\bar K^0,\,\eta\eta)$ for $r=f_0(980)$; $\Pi^{ab}_r(s)$ stands for
the diagonal matrix element of the polarization operator of the
resonance $r$ corresponding to the contribution of the $ab$
intermediate state \cite{ADS80}.

At $s>(m_a+m_b)^2$,
\begin{eqnarray}\label{Eq14}\Pi^{ab}_{r}(s)=\frac{g^2_{r
ab}}{16\pi} \left[\frac{m_{ab}^{(+)}m_{ab}^{(-)}}{\pi
s}\ln\frac{m_b}{m_a}+\rho_{ab}(s)\right.\ \nonumber\\
\left.\times\left(i-\frac{1}{\pi}\,\ln\frac{\sqrt{s-m_{ab}^{(-)
\,2}}+\sqrt{s-m_{ab}^{(+)\,2}}}{\sqrt{s-m_{ab}^{(-)\,2}}-\sqrt{s
-m_{ab}^{(+)\,2}}}\right)\right],\end{eqnarray} where $g_{rab}$ is
the coupling constant of $r$ with $ab$, $\rho_{ab}(s)$\,=\,$
\sqrt{s-m_{ab}^{(+)\,2}} \,\sqrt{s-m_{ab}^{(-)\,2}}\,/s$,
$m_{ab}^{(\pm)}$\,=\,$m_a\pm m_b$, and $m_a\geq m_b$; $\mbox{Im}
\,\Pi^{ab}_r(s)=\sqrt{s} \Gamma_{r\to ab}(s)=(g^2_{r
ab}/16\pi)\rho_{ab}(s)$. At $m_{ab}^{(-)\,2}<s<m_{ab}^{(+)\,2}$
\begin{eqnarray}\label{Eq15}\Pi^{ab}_{r}(s)=\frac{g^2_{r
ab}}{16\pi} \left[\frac{m_{ab}^{(+)}m_{ab}^{(-)}}{\pi
s}\ln\frac{m_b}{m_a}\right.\nonumber\\
\left.-\rho_{ab}(s)\left(1-\frac{2}{\pi}\arctan\frac{\sqrt{
m_{ab}^{(+)\,2}-s}}{\sqrt{s-m_{ab}^{(-)\,2}}}\right)\right],
\end{eqnarray}
where  $\rho_{ab}(s)$\,=\,$\sqrt{m_{ab}^{(+)\,2}-s}
\,\sqrt{s-m_{ab}^{(-)\,2}}\,/s$. At $s\leq m_{ab}^{(-)\,2}$
\begin{eqnarray}\label{Eq16}\Pi^{ab}_{r}(s)=\frac{g^2_{r
ab}}{16\pi} \left[\frac{m_{ab}^{(+)}m_{ab}^{(-)}}{\pi
s}\ln\frac{m_b}{m_a}\right.\ \ \nonumber\\
\left.+\rho_{ab}(s)\frac{1}{\pi}\,\ln\frac{
\sqrt{m_{ab}^{(+)\,2}-s}+\sqrt{m_{ab}^{(-)\,2}-s}}
{\sqrt{m_{ab}^{(+)\,2}-s}-\sqrt{m_{ab}^{(-)\,2}-s}}\right],
\end{eqnarray}
where $\rho_{ab}(s)$\,=\,$\sqrt{m_{ab}^{(+)\,2}-s}\,\sqrt{
m_{ab}^{(-)\,2}-s}\,/s$.

The propagators $1/D_{a^0_0}(s)$ and $1/D_{f_0}(s)$ constructed with
taking into account the finite width corrections [see
Eqs.~(\ref{Eq13})--(\ref{Eq16})] satisfy the K\"{a}ll\'{e}n-Lehman
representation in the wide domain of coupling constants of the
scalar mesons with two-particle states and, due to this fact,
provide the normalization of the total decay probability to unity:
$\sum_{ab}BR(r\to ab)=1$ \cite{AKi04}.

Here we use the numerical estimates of the coupling constants
$g^2_{f_0 ab}/(16\pi)$ and $g^2_{a^0_0ab}/(16\pi)$ obtained in Ref.
\cite{AKS16}
\begin{eqnarray}\label{Eq17}
\frac{g^2_{f_0\pi\pi}}{16\pi}\equiv\frac{3}{2}\frac{g^2_{f_0\pi^+\pi^-}}
{16\pi}=0.098\mbox{\ GeV}^2,\\ \label{Eq18} \frac{g^2_{f_0 K\bar
K}}{16\pi}\equiv2\frac{g^2_{f_0 K^+K^-}}{16\pi}=0.4\mbox{\ GeV}^2,
\\ \label{Eq19} \frac{g^2_{a^0_0\eta\pi^0}}{16 \pi}=0.2\mbox{\ GeV}^2,
\quad\quad\  \\ \label{Eq20} \frac{g^2_{a^0_0 K\bar K}}{16\pi}
\equiv2\frac{g^2_{a^0_0 K^+K^-}}{16\pi}=0.5\mbox{\ GeV}^2.
\end{eqnarray}
As in Ref. \cite{AKS16}, we fix $m_{a^0_0}=0.985$ GeV, $m_{f_0}=
0.985 $ GeV and set $g^2_{a^0_0\eta'\pi^0}=g^2_{a^0_0\eta\pi^0}$ and
$g^2_{f_0\eta\eta}=g^2_{f_0 K^+K^-}$ by the $q^2\bar q^2$ model,
see, e.g., Refs. \cite{ADS81,AI89}.



\begin{thebibliography}{99}
\bibitem{FN2}  The sharp jump of the phase suggests an idea to study the $a^0_0(980)-f_0(980)$ mixing in
               polarization phenomena, see Refs. \cite{AS04a,AS04b}. These effects are still in waiting for
               their experimentalists.
\bibitem{FN2a} The points with error bars in this plot correspond to the square of the $\pi^+\pi^-$ $S$ wave
               amplitude given numerically in Table II in Ref. \cite{BaBar09}. Errors are statistical only.
\bibitem{FN3} Nevertheless, we must note that if the $f_0(980)$
is produced as a solitary resonance without any background in the
$\pi^+\pi^-$ and $K^+K^-$ channels, then the CLEO data \cite{CLEO13}
on the branching ratio $BR(D_s^+\to\pi^+\pi^-\pi^+)=(1.11\pm
0.04\pm0.04)\%$ (see also Refs. \cite{BaBar09,PDG16}) contradict the
BaBar \cite{BaBar11} and CLEO \cite{CLEO09} assumption of the
$f_0(980)$ dominance in the decay $D^+_s\to f_0(980)\pi^+\to K^+K^-
\pi^+$. Remember that the PDG average $BR(D^+_s\to f_0(980)\pi^+\to
K^+K^-\pi^+)=(1.15\pm0.32)\%$ [see Eq. (\ref{Eq1})]. The situation
requires clarification. By our estimate $BR(D^+_s\to f_0(980)\pi^+
\to\pi^+\pi^-\pi^+)\approx3.4\%$ is in agreement with Eq.
(\ref{Eq1}) and the curve in Fig. 2. Let us emphasize once again
that the detection of the $a^0_0(980)-f_0(980)$ mixing can provide a
unique opportunity to clarify the puzzling mechanisms of the
$f_0(980)$ and $a^0_0(980)$ production and decay in the three-body
hadronic decays of the $D^+_s$ meson. The branching ratio of the
solitary $f_0(980)$ resonance decay into $\pi^+\pi^-$ is larger than
into $K^+K^-$ because of the phase space factor of the $K^+K^-$
system that suppresses the $f_0(980)\to K^+K^-$ decay near the
threshold.
\bibitem{FN4} Note that the direct production of the
$(a_0(980)\pi)^+$ system via the $D^+_s\to W^+\to\left(a^+_0(980)
\pi^0+a^0_0(980)\pi^+\right)$ transition (the annihilation
amplitude) is impossible, i.e., the matrix element of the isovector
axial vector current $\langle0| j_\mu^A|(a_0(980)\pi)^+\rangle=0$.
The matter is that the isovector axial vector current has the
negative $G$ parity in Standard Model, whereas the $G$ parity of the
$(a^0(980)\pi)^+$ system is positive. The final $(a_0(980)\pi)^+$
system is produced in the weak $D^+_s$ decay in the state with
isospin $I=1$. This implies that the effective interaction
Lagrangian responsible for the transition
$D^+_s\to(a_0(980)\pi)^+_{I=1}$ is proportional to $(a^+_0(980)\pi^0
-a^0_0(980) \pi^+)$. Thus, the amplitude $A_{a_0\pi}$, like
$A_{\eta\rho^+}$, is antisymmetric with respect to permutation of
the $s$ and $t$ variables.
\end{thebibliography}
\end{document}